\documentclass[amsmath,twocolumn]{revtex4}

\usepackage{graphicx}
\usepackage{tikz}

\usepackage{lipsum}
\usepackage{amssymb}
\begin{document}
\title{Scattering by finite periodic $\mathcal{PT}$-symmetric structures} 
\author{V.~Achilleos}
\author{Y.~Aur\'{e}gan}
\author{V.~Pagneux}
\affiliation{ Laboratoire d'Acoustique de l'Universit\'e du Maine, UMR CNRS 6613 Av. O. Messiaen, F-72085 LE MANS Cedex 9, France}

\begin{abstract}
In this work, we study the transmission properties of one dimensional finite periodic systems with $\mathcal{PT}$-symmetry.
A simple closed form expression is obtained for the total transmittance  from a lattice of N cells, that allows us
to describe the transmission minima (maxima) when the system is in the $\mathcal{PT}$-unbroken (broken) phase. Utilizing this expression, we  provide the necessary conditions, \textit{independent}
of the number of cells, for the occurrence of a CPA-laser for any finite  $\mathcal{PT}$-symmetric periodic potential.
Under these conditions, we provide a recipe for building finite periodic structures with near perfect absorption 
and extremely large amplification. 

\end{abstract}

\maketitle

%

Studying the interplay between losses and gain in wave propagation
has recently attracted considerable attention, stimulated by the discovery 
of  $\mathcal{PT}$-symmetric~\cite{Bender1998} systems. 
Theoretical efforts where initially focused on  extending Hermitian quantum theories~\cite{Bender1999a,Bender2007} 
using non-Hermitian $\mathcal{PT}$-symmetric Hamiltonians with real spectra. 
It has been however realized that the notion of $\mathcal{PT}$-symmetry and the corresponding
phenomena can be readily extended to other physical systems.
Experiments emulating $\mathcal{PT}$-symmetric Hamiltonians 
are now performed in diverse physical settings including optical waveguides~\cite{Guo2009,Ruter2010},
micro-ring resonators~\cite{Peng14a,Peng14b,Chang,Feng14,Hodaei},
audible acoustics~\cite{Fleury2015, famousPT}, optomechanics~\cite{Lu2015},
spin-waves~\cite{Peng2016}, or  atomic systems~\cite{Zhang2016}
The interest in such systems is motivated by the 
extraordinary wave properties, especially around exceptional points~\cite{Tref2005,Moiseyev2011} 
which are otherwise unattainable in Hermitian systems
including, unidirectional propagation~\cite{Lin2011,Feng2013,Longhiuni}, enhanced sensitivity~\cite{Liu2016} and coherent perfect 
absorbers-lasers (CPA-lasers)~\cite{Longhi2010-a,Chong2011}.
The CPA-laser, especially in optics, is of great technological interest since it may lead
to devices which are multifunctional, acting on the same time as absorbers, lasers or modulators.
Experimental observation of the CPA-laser was first realized in an electronic circuit analogue~\cite{Kottos2012}
and it has been very recently reported in an optical setting~\cite{CPAlaser2016}. 
 
$\mathcal{PT}$-symmetry requires a delicate exact balance between gain and loss.
 which makes  the experimental observation of the $PT$-phase transition as well as the CPA-laser  a very challenging task.
One way to bypass the difficulties is to try and build  elemental two-component systems
and fine-tune loss/gain the obtain exact $\mathcal{PT}$-symmetry.
On the other hand, periodic structures with $\mathcal{PT}$-symmetry are 
of special interest~\cite{Bender1999,Shin2004,Musslimani2008,Midya2010,Graefe2011}, since they have been shown to 
have interesting properties including  unusual band structure and Bloch 
oscillations~\cite{makris2008,Longhi2009,Longhi2010-b,makris2010,Regensburger2012}.
They offer a unique opportunity for generating novel devices, and constitute
a promising setting to overcome the consequences of losses in many applications
including the growing field of metamaterials. 

Although many studies exist for the case of  infinite periodic $\mathcal{PT}$-symmetric
systems, the case of scattering in a finite periodic systems composed of $N$ number of cells has been
less investigated. In many cases (see for example Ref.~\cite{Lin2011} and 
the experimental works~\cite{Feng2012,CPAlaser2016,metawaveguide})
the finite system is studied using a coupled mode theory around the Bragg points
and an approximate transfer matrix describes the relevant phenomena.

It was recently shown in Ref.~\cite{Tsironis2016} that asymmetric transmission resonances (ATR)~\cite{Ge2012}
(exceptional points of the scattering matrix) and the CPA-laser points of a finite number 
$\mathcal{PT}$-symmetric dielectric layers highly depend on the number of cells.
A finite lattice of N-quantum $\mathcal{PT}$-symmetric scatterers was studied 
in Ref.~\cite{Yu2016} focusing on the difference between parallel and in-series 
coupling of the scatterers, but also relating the singular points 
of the scattering matrix the unit cell and the N-cell structure. Even more
recently the singular value spectrum of a finite periodic electromagnetic
structure was studied in Ref.~\cite{LiGe2017}, focusing on the CPA-laser point.

The purpose of this article is to give simple closed form expressions, describing
the transmission and the CPA-laser points, from an arbitrary one dimensional finite periodic 
$\mathcal{PT}$-symmetric scatterer. The expression for the total transmission from the finite system
is given in  Eq.~(\ref{TN}), which depends on the unit cell transmission, the Bloch phase and the total number of cells. 
From this expression we can deduce the number of transmission resonances, the ATRs and the CPA-laser points.
Furthermore we obtain an envelope function, Eq.~(\ref{avg}), which captures the minima of transmission 
in the $\mathcal{PT}$-unbroken phase or  the maxima in the $\mathcal{PT}$- broken phase.  
Using this simple form we obtain the necessary conditions for a CPA-laser,
in any finite periodic $\mathcal{PT}$-symmetric potential in 1D, independently of the number of cells (length).
These conditions are given in Eq.~(\ref{conditions}). The exact CPA-laser which depends
on the number of cells $N$ is also found here analytically. Finally we show that when the 
necessary requirements obtained by the envelope function are met, near perfect absorption 
and extremely strong amplification at the same frequency can be obtained even away from the CPA-laser.

\begin{figure}[tbp!]
\includegraphics[width=8.5cm]{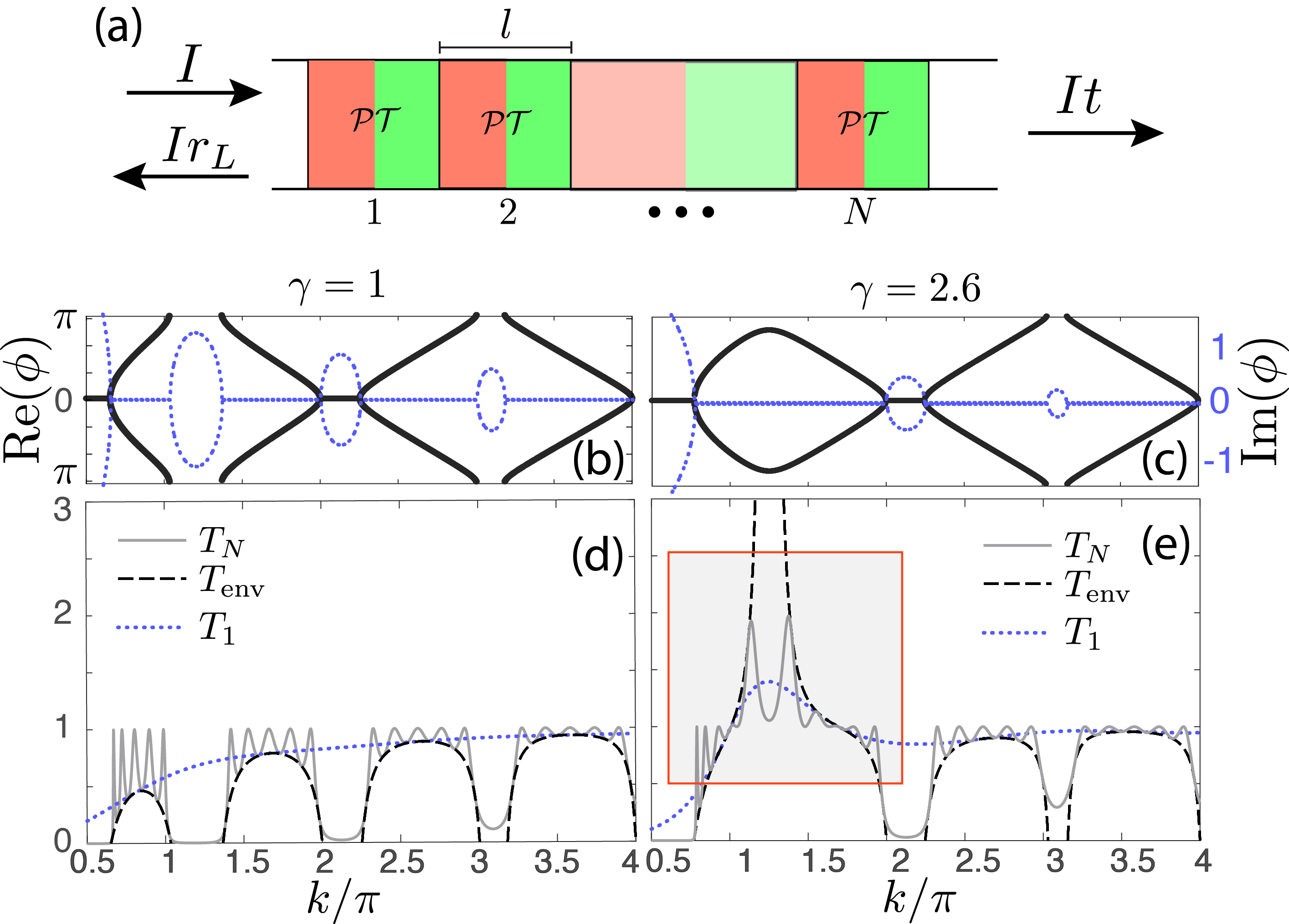}
\caption{Panel (a): schematic of the unit cell and the finite periodic potential of N cells. Arrows indicate an incoming
wave $I$ form the left and the corresponding transmitted and reflected waves.
Panels (b) and (c): the real and imaginary part of the Bloch phase $\phi$ as a function of wavenumber $k/\pi$
obtained for the unit cell of Eq.~(\ref{delta}).
Panels (e) and (d) the transmittance $T_N$ through a finite  lattice of $N=6$ cells as a function of $k/\pi$.
Left and right columns correspond to gain/loss parameter $\gamma=1$ and $\gamma=2.6$ respectively, while
in all cases $\lambda=6$ and $l=1$.}
\label{delta1} 
\end{figure}

In the following, we study one dimensional scattering systems satisfying the stationary  Schr\"{o}dinger equation 
\begin{equation}
\psi''+(k^2-V(x))\psi=0,
\label{helm1}
\end{equation}
or the Helmholtz equation relevant to optical Bragg gratings
\begin{equation}
\psi''+k^2n^2(x)\psi=0
\label{helm2}
\end{equation}
where $\psi(x)$ is the wave field and primes denote derivatives with respect to $x$.
The  $\mathcal{PT}$-symmetry of the potential is established when $V(x)=V^*(-x)$, where star denotes complex conjugation,
while the normalized refractive index $n(x)$ also is  $\mathcal{PT}$-symmetric when $n(x)=n^*(-x)$.
We study space periodic potentials of period $l$ satisfying $V(x+l)=V(x)$ and $n(x+l)=n(x)$
as shown in Fig.~\ref{delta1} (a).
For both equations, the scattering matrix for the unit cell with length $l$ has the form 
\begin{eqnarray}
\mathbf{S_1}=\left(
\begin{array}{cc} r_L&t\\t&r_R\end{array} \right),
\end{eqnarray}
where  $t$,  $r_L$ and $r_R$ are the transmission, reflection from left and right coefficients respectively.
The transfer matrix of the unit cell defined in the region  $-l/2\leq x \leq l/2$,
 is $\mathcal{PT}$-symmetric and can be written  as:
\begin{eqnarray}
\mathbf{M_1}=\left(
\begin{array}{cc} 1/t^*&r_R/t \\  -r_L/t &1/t	\end{array} \right).
\label{M1}
\end{eqnarray}
$\mathbf{M_1}$  has a unitary determinant which leads to the following
relation 
%
$r_Lr_R=t^2(1-T_1^{-1})$
%
and subsequently to the following "conservation law"
%
$|T_1-1|=\sqrt{R^{(1)}_LR^{(1)}_R}$.
%
Here $T_1=|t|^2$ is the total transmittance from the unit cell and $R^{(1)}_{L,R}=|r_{L,R}|^2$ the total 
reflectances of the unit cell from left and right. Since the system does not conserve time reversal
symmetry in general $R^{(1)}_{L}\ne R^{(1)}_{R}$.

Depending on the parameter values of the unit cell scatterer, the transmittance may either be $T_1<1$ or $T_1>1$.
Besides, the eigenvalues of $\mathbf{M}_1$ can be written  as $\lambda_{1,2}=\exp[i\phi]$ resulting in
\begin{eqnarray}
\cos\phi=Re(1/t).
\label{bloch}
\end{eqnarray}
In the case of an infinite periodic potential the phase $\phi$ corresponds to the Bloch phase.
\paragraph*{Transmittance---}
We now focus on the scattering from a finite periodic structure composed by $N$ cells.
Thus,  both the potential  $V(x)$ or the refractive index $n(x)$ are $\mathcal{PT}$-symmetric in a 
region $-L/2<x<L/2$ where the total length of the scatterrer is  $L=Nl$.
Using the classical Chebyshev identity we can write the transfer matrix for a periodic
potential composed by $N$ cells~\cite{Tsironis2016,LiGe2017}, which has the form 
%
\begin{eqnarray}
%
\mathbf{M}_N=&\left(
\begin{array}{cc}  \frac{1}{t^*}\frac{\sin(N\phi)}{\sin\phi}-\frac{\sin(N-1\phi)}{\sin\phi} & 
\frac{r_R}{t}\frac{\sin(N\phi)}{\sin\phi}\\ -\frac{r_L}{t}\frac{\sin(N\phi)}{\sin\phi}&\frac{1}{t}\frac{\sin(N\phi)}{\sin\phi}-\frac{\sin(N-1\phi)}{\sin\phi}\end{array} \right) 
\nonumber \\
\label{MN}
\end{eqnarray}
Since the potential is still $\mathcal{PT}$-symmetric, the conservation relation 
is now generalized to
$|T_N-1|=\sqrt{R^{(N)}_LR^{(N)}_R}.$
Using the fact that $\mathbf{M}_N$ also has a unitary determinant, and that  it is also $\mathcal{PT}$-symmetric,
from Eq.~(\ref{MN}) we obtain the total transmission from $N$ cells as
\begin{align}
\frac{1}{T_N}=
1+(\frac{1}{T_1}-1)\frac{\sin^2(N\phi)}{\sin^2\phi}.
\label{TN}
\end{align}
The transmission is obtained  as a function of the
phase $\phi$, the transmission of the unit cell $T_1$ and the number of cells $N$.
It is very interesting to note that  Eq.~(\ref{TN}) has the same form as the transmission through 
a periodic scatterer without loss or gain~\cite{sprung}.  
In  Ref.~\cite{sprung} the expression of Eq.~(\ref{TN}) is derived for the conservative system
exploiting  the fact that the scattering matrix $\mathbf{S}_1$ is unitary. On the contrary, for the 
$\mathcal{PT}$-symmetric case we use the "generalized unitarity relation" (see for example Ref.~\cite{Ge2012}).
To our knowledge this simple and useful expression has not been established in the literature  for $\mathcal{PT}$-symmetric
periodic potentials.

The simplicity of Eq.~(\ref{TN}) allows us to identify most of the characteristics of the scattering problem
with simple arguments, and as we present below, it also permits to extract the necessary conditions
for CPA-laser in any 1D finite periodic system.
To illustrate the use of  Eq.~(\ref{TN})  we study 
a $\mathcal{PT}$-symmetric periodic potential composed by  $\delta$ barriers for Eq.~(\ref{helm1}).
The unit cell is described by the following potential
 \begin{eqnarray}
V(x)=\lambda\delta(x)-i\gamma\left(\delta(x-\frac{l}{4})-\delta(x+\frac{l}{4})\right),
\label{delta}
 \end{eqnarray}
defined in the region $-l/2<x<l/2$ with  a total length $l$. 

In the unbroken phase ($T_1<1$), the finite periodic $\mathcal{PT}$-symmetric has a similar
behavior as in the conservative case (no loss and no gain), and in order to be self contained we
review these properties~\cite{sprung}. The transmission $T_N$, acquires $N-1$ transmission resonances (TR) within each 
propagating band,  corresponding to  the values of  $N\phi=n\pi$ with $n=1,2\ldots N-1$.  
An example of the transmittance from  $N=6$ cells of the form of Eq.~(\ref{delta}), in the $\mathcal{PT}$-unbroken
phase,  is shown in Fig.~\ref{delta1} (d), featuring $5$ TR in each band. 
Note that, according to Eq.~(\ref{TN}), if the unit cell admits additional 
resonances ($T_1=1$) these have to be added to the $N-1$ stemming from the periodicity.
Importantly, the aforementioned reasoning for counting the TRs is based on the fact that the Bloch phase $\phi$
inside each transmitting band is  a \textit{monotonous} function describing the region  $\phi\in[0,\pi]$, for the
unbroken phase.  The Bloch phase of the unit cell~(\ref{delta}) with $\gamma=1$, is shown in Fig.~\ref{delta1} (b). 
%
%
Additionally, the minima of  transmission inside each band, appear when  $N\phi=n\pi/2$.

In the broken phase ($T_1>1$), according to Eq.~(\ref{TN}), the transmission 
changes drastically. Although there exist  TRs at $N\phi=n\pi$, now there also 
peaks of superradiant transmission with $T_N>1$ at $N\phi=n\pi/2$. 
Furthermore, for values of  $k$, between the beginning of the first band and $k=2\pi$ which is now 
a region of propagation, $\phi$ is no longer monotonic. In fact, it starts from $\phi=0$ and returns to $0$
without passing from $\phi=\pi$ as shown in Fig.~\ref{delta1} (c).
An important consequence of the trajectory of $\phi$ is that the number of transmission resonances 
and peaks is now system dependent, and does not follow a prescribed rule as in the conservative system.
\begin{figure}[tbp!]
\includegraphics[width=8.5cm]{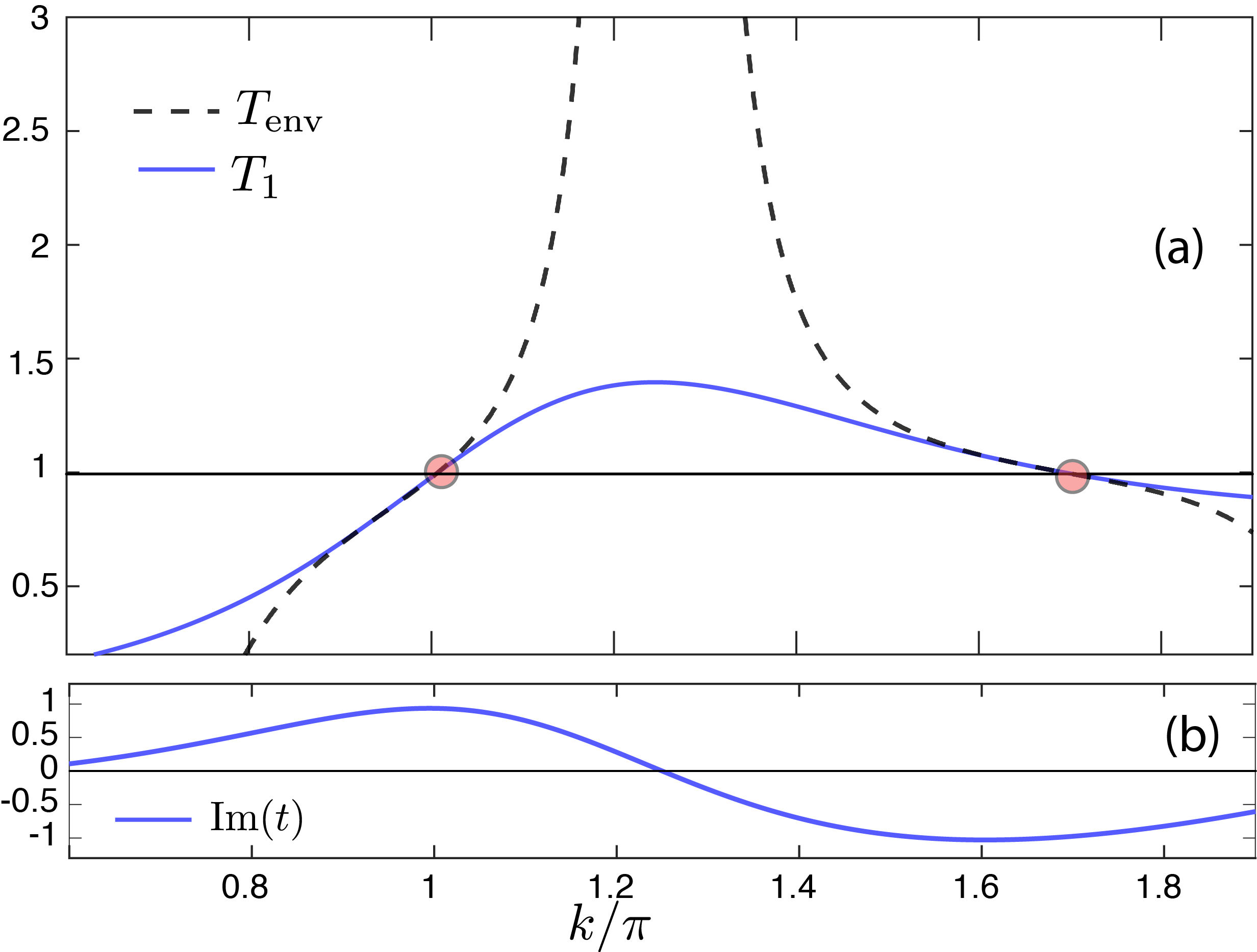}
\caption{Panel (a): The transmittance $T_1$ (dashed) and the envelope function $T_{\rm env}$ (solid)
around the $\mathcal{PT}$-broken parametric region where  $T_1>1$ for $\gamma=2.6$.
 Panel (b): The imaginary part of the unit cell transmission coefficient $t$ in the same region. }
\label{envelopecond} 
\end{figure}
\paragraph*{Envelope transmission function and CPA-laser condition--- }
A  simple envelope function~\cite{sprung}, which is \textit{independent} of the number of cells $N$
can be obtained by considering  the points where $\sin(N\phi)=1$ which,
using Eq.~(\ref{TN}), leads to 
\begin{align}
\frac{1}{T_{\rm env}}=1-\frac{(1-\frac{1}{T_1})}{\sin^2\phi}.
\label{avg}
\end{align}
We first note that when unit cell is in the unbroken phase ($T_1<1$), the envelope function 
$T_{\rm env}$  describes the minima of the total transmission $T_N$~\cite{sprung}
in the propagating bands, as is the case for a conservative system. 
This is shown in Fig.~\ref{delta1} (d) by the dashed  black line. Even for a lattice as small as with six cells,
 this $N$ independent function captures the envelope of the minima efficiently.
On the other hand, when the unit cell is in the broken phase ($T_1>1$) the function  $T_{\rm env}$ now
describes the maxima of $T_N$ as shown in  Fig.~\ref{delta1} (e).
Importantly, as it is shown in Fig.~\ref{delta1} (e), the envelope function $T_{\rm env}$ 
becomes infinite in the region where $T_1>1$ when, according to Eq.~(\ref{avg}),
 $\sin^2\phi=(1-1/T_1)$. Such an infinite transmission in a $\mathcal{PT}$-symmetric 
potential is known to correspond to a CPA-laser point~\cite{Longhi2010-a,Chong2011}.
The CPA-laser corresponds to the case where one eigenvalue of the total scattering matrix
goes to infinity (laser) while the other vanishes (absorber).
Due to this property, when the system is found to exhibit huge transmission it also expected to
act as  near perfect absorber at the same frequency and for the same parameters.

Using Eq.~(\ref{bloch}) we  find the necessary condition, which is $N$ \textit{independent}
, in order for a finite periodic system to feature a CPA-laser
\begin{eqnarray}
|t|>1,\quad {\rm and} \;\; {\rm Im} (t)=0.
\label{conditions}
\end{eqnarray}
The condition of Eq.~(\ref{conditions}) depends solely on $t$, is independent of the number  $N$ 
of scatterers, and is  necessary in order to obtain a CPA-laser in the periodic structure.
In Fig.~\ref{envelopecond} (a), we show the single cell transmission $T_1$ (solid line) 
corresponding the same values as in Fig.~\ref{delta1} (e), around the superradiant region. 
With the dashed line, we plot $T_{\rm env}$ which diverges at the value of the  point 
where the imaginary part of $t$ [see Fig.~\ref{envelopecond} (b)] crosses zero.
%

%
For the finite system with $N$ cells, Eq.~(\ref{conditions})
designates the parametric region where the CPA-laser is able to appear.
The  exact condition for a CPA-laser  requires $N=(2n+1)\pi/2\phi$
with arbitrary $n$, and if satisfied it will lead to infinite transmittance. 
In many realistic applications, after a sufficiently high amplification of the wave field
nonlinear effects become important and the linear theory is no longer valid.
It  thus  practical, to propose  structures able to nearly absorb and efficiently amplify incoming waves
but with a finite rate.	

\begin{figure}[tbp!]
\includegraphics[width=8.5cm]{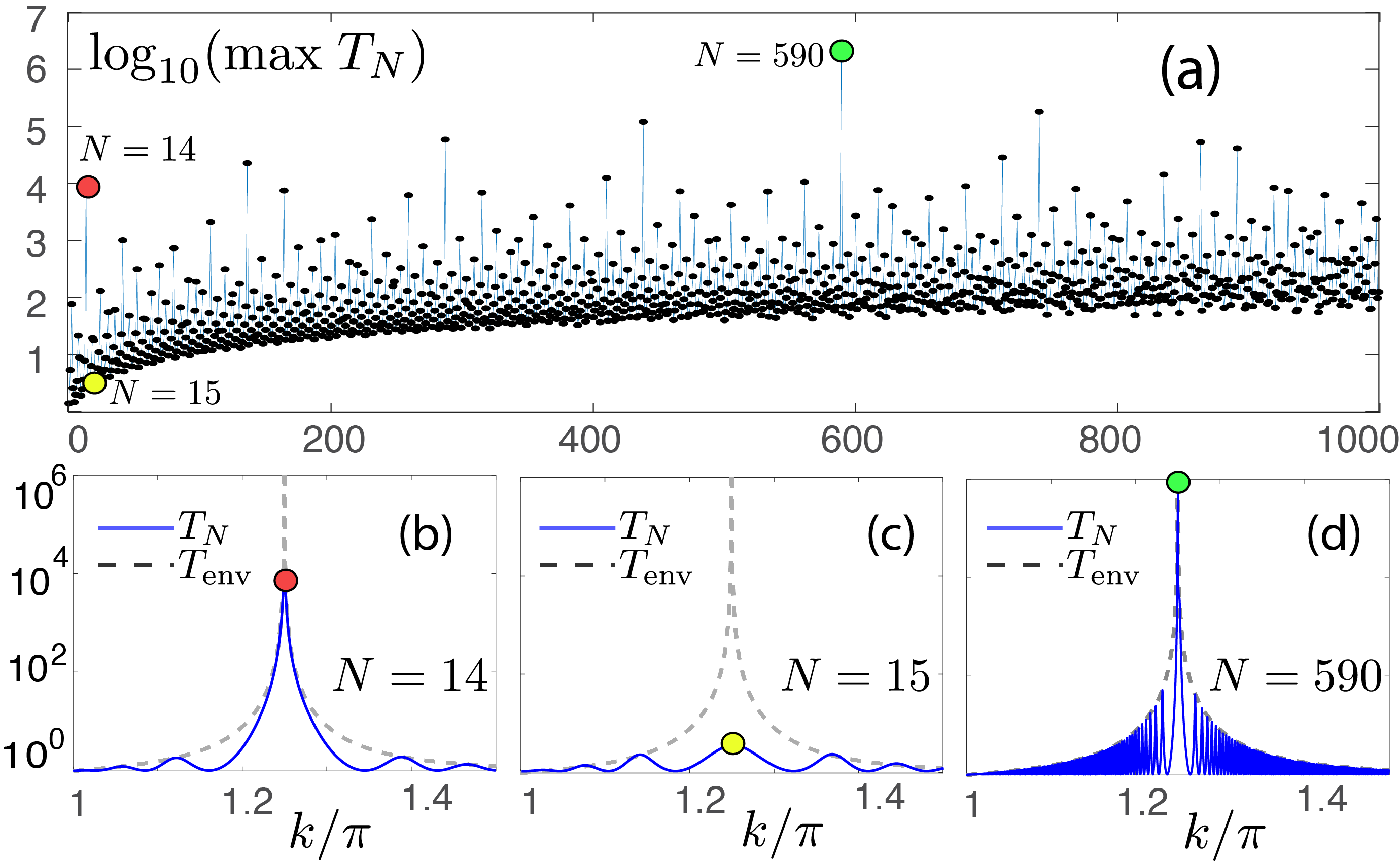}
\caption{Panel (a): The maximum of the logarithm of  $T_N$ as a function of the total number of cells N.
Panels (b), (c) and (d): The transmittance $T_N$ for three different cases  as indicated by the number of cells.
Parameters used are the same as in the right column of Fig.~\ref{delta1} }
\label{increasingN} 
\end{figure}

Here we like to stress the fact that having a single cell satisfying Eq.~\ref{conditions}, allows one to then 
vary the number of cells and to obtain huge amplification/absorption.
For the case of $\gamma=2.6$ used in the previous examples, we plot in Fig.~\ref{increasingN}(a) the maximum
of transmittance as a function of the number of cells. We observe that the maximum transmission varies 
significantly even with a small change in the number of cells. The bottom part of the curve in Fig.~\ref{increasingN}(a),
appears to be saturating for large lattices. Importantly, we observe different peaks, within the range of $N$ plotted here,
which correspond to an amplification of the wave up to $10^7$ times. 

Examples of the transmission for three different number of cells are shown in Fig.~\ref{increasingN} (b), (c) and (d).
In the case of Fig.~\ref{increasingN} (b), we have chosen $N=14$ since it appears to have a 
huge maximum transmission, in the case of only a few unit cells. In contrast just by adding 2 cells,
the transmission in Fig.~\ref{increasingN} (c) has a maximum less than 10.
Furthermore, using panel (a), we can choose the maximum possible transmission in this range of cells,
which appears at $N=590$, and the corresponding $T_N$ is shown in  Fig.~\ref{increasingN} (d), reaching
upt to an amplification of $10^6$.

It is interesting to note that the unit cell crosses from the unbroken to the broken phase through a TR of
the unit cell with $T_1=1$ which is also a TR of the finite periodic structure. Two such TRs are indicated
with (red) circles in Fig.~\ref{envelopecond} (a). According to the modified conservation law for $\mathcal{PT}$-symmetric systems
[see below Eq.~(\ref{M1})], here there is a totally asymmetric reflection since  either $R_L^{(1)}$ or $R_R^{(1)}$ is 
zero (except if there is an accidental zero for both reflections). This transmission resonance was discussed in Ref.~\cite{Ge2012} and corresponds to an exceptional 
point of the scattering matrix of the unit cell. Due to the form of the transfer matrix in Eq.~(\ref{MN}), the reflections from 
the finite lattice $R_{L,R}^{(N)}$  are analogous to the ones of the unit cell, and thus this ATRs also appear
for the periodic structure.

\begin{figure}[tbp!]
\includegraphics[width=8.5cm]{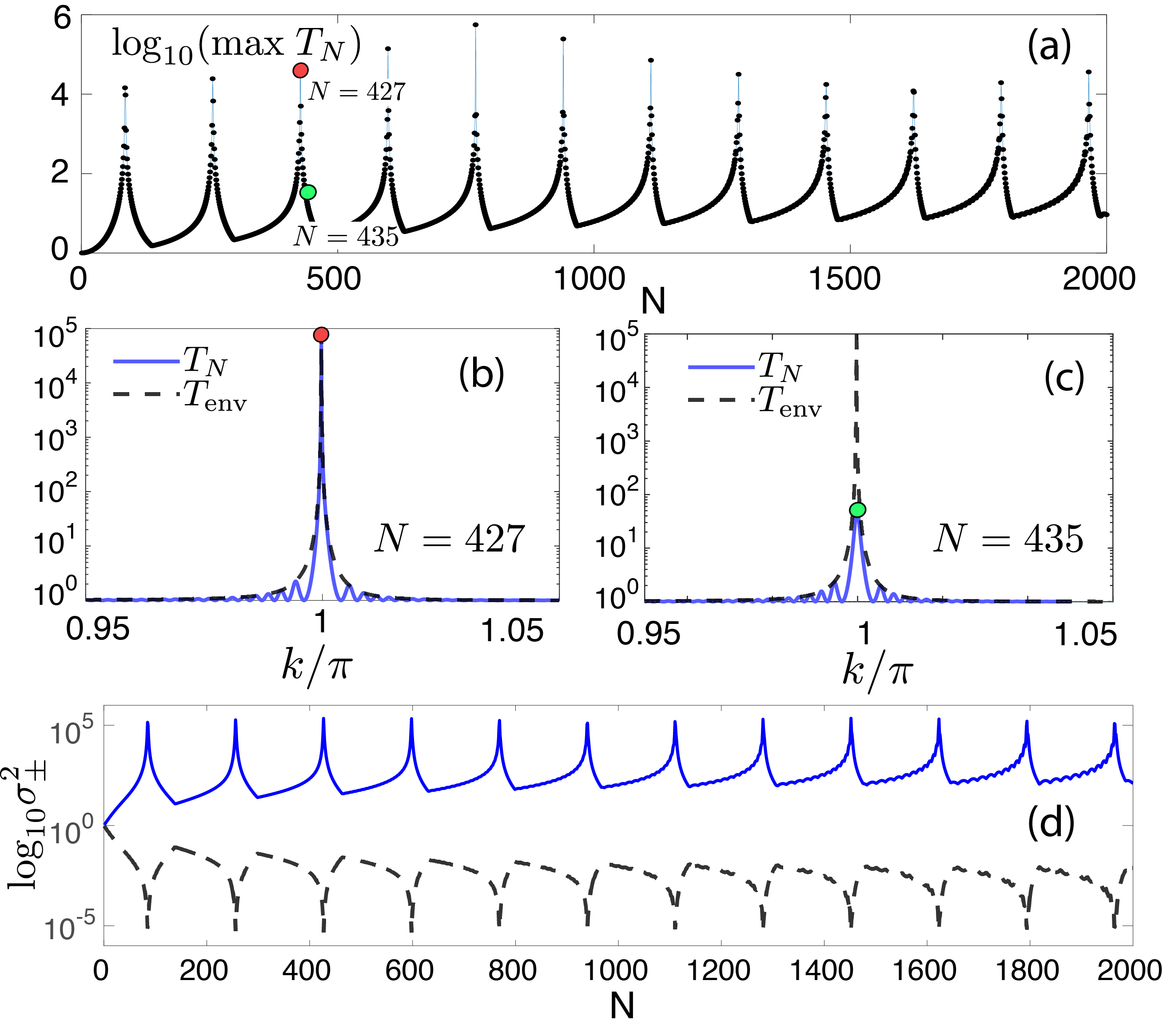}
\caption{Panel (a), (b) and (c): same as the corresponding panels  of Fig.~\ref{increasingN} 
but for the case of the electromagnetic slabs with parameters $n_1=0.02$, and $\gamma=1.1$. 
Panel (d) shows the logarithm of the singular values $\sigma_\pm^2 $ of the scattering matrix
as a function of the number of cells.}
\label{gratings} 
\end{figure}

\paragraph*{Electromagnetic gratings---}
Now, we apply the aforementioned results in a system that has been extensively studied
in the context of $\mathcal{PT}$-symmetry and periodicity i.e. a lattice of electromagnetic gratings or slabs. 
This is one of the few systems which has exhibited a CPA-laser experimentally in optics~\cite{Zhang2016}
and it is  very promising for for next-generation photonic integrated circuits. 
For such a system the transverse electric field satisfies  Eq.~(\ref{helm2}) with a refractive index $n(x)$ given by
\begin{eqnarray}
n(x)=1+n_1\left[{\rm sgn}\left(\cos 2l x\right)+i\gamma\;{\rm sgn}\left(\sin 2 l x\right)\right].
\label{Vslabs}
\end{eqnarray}
where we consider a piecewise constant refractive index in each slab of length $l/4$.
For this particular distribution of the refractive index, it is  known~\cite{Musslimani2008,Midya2010,Graefe2011} that 
the  $PT$-broken phase appears for $\gamma\geqslant 1$.
 The system is usually analyzed by the use of a coupled mode theory, valid 
around the Bragg frequency. The length of the total finite lattice, and thus the number of cells,
is then chosen using the transfer matrix stemming from the coupled mode theory.

Since for the unbroken phase we have shown that the transmission properties are similar to the 
conservative case, here we directly focus on the  $\mathcal{PT}$-broken phase with  $\gamma>1$.
In particular by choosing $n_1=0.02$ and $\gamma=1.1$ we identify the region where 
the conditions of Eq.~(\ref{conditions}) are  satisfied.  In fact the CPA-laser for this settings is found
close to the Bragg wavenumber as it is expected~\cite{Musslimani2008, Zhang2016,LiGe2017}.
Fig.~\ref{gratings} (a) shows the logarithm of the maximum transmission $T_N$ as a function of the number of cells
N. The dependence on the number of cells shows a structured pattern in contrast to the one of Fig.~\ref{increasingN}.
However, also in this case the minimum of ${\rm log}({\rm max} T_N)$ appears to saturate to a limiting value
for large N.  Its is again shown that  by satisfying Eq.~(\ref{conditions}),
we can  change the number of cells in order to obtain configurations with huge amplification/absorption.
Two examples, of the transmission are shown in Fig.~\ref{gratings} (b) and (c), illustrating again how the number
of cells can be used in order to obtain large amplification.

In order to describe  quantitatively,  both amplification and absorption  for the case of the electromagnetic 
slabs, we perform a singular value  decomposition of the scattering matrix from $N$ cells. 
The singular values $\sigma_{\pm}$ are both real  and non-negative, and more importantly they
 are connected through $\sigma_+\sigma_-=1$ for a 1D  $\mathcal{PT}$-symmetric system~\cite{LiGe2017}. 

For a given incoming wave of the form 
$\vec{X}= [\psi^+,\psi^-]^T$ where $\psi^{\pm}$ are the forward and backward propagating
waves, one can define the output to input ratio as $\Theta=\|\mathbf{S}\vec{X}\|/\|\vec{X}\|$.
Then the two singular values correspond to the $\sigma_-={\rm min}[\Theta$] and
$\sigma_+={\rm max}[\Theta$], and quantify the ability of the scatterer to absorb or amplify an incoming 
wave $\vec{X}$.
At the CPA-laser point, the singular values become $\sigma_-=0$ and $\sigma_+\rightarrow\infty$.
Here to illustrate the absorptive and amplifying  properties of the system Fig.~\ref{gratings} (d), 
we plot $\sigma_{\pm}$ as  a function of $N$. It is found that by satisfying the conditions of
Eq.~(\ref{conditions}) and varying the number of cells, one is able to obtain a high contrast of absorption
and amplification at the same frequency. The dependence on the number of cells is also very sharp in this
case.

In summary, we studied finite periodic $\mathcal{PT}$-symmetric scatterers in one dimension. 
We have obtained closed form expression for the transmission from a finite system of $N$ cells,
as a function of the single cell transmission, the Bloch phase and $N$. 
A simple envelope function \textit{independent} of the number of cells was obtained, describing 
the minima (in the $\mathcal{PT}$-unbroken phase) or the maxima (in the $\mathcal{PT}$-unbroken phase)
of the total transmission. Using this function we find the necessary conditions for CPA-laser, for an arbitrary 1D
periodic 1D system, depending only on the unit cell.
Although the exact CPA-laser point depends on its total length, here
we show that when the necessary conditions, obtained for the unit cell, are met by varying the number of cells, one can 
achieve huge amplification and high absorption at the same frequency.

\end{document}